\documentclass[
  twocolumn,
  prl,
  showpacs,
  amsmath,
  amssymb,
  superscriptaddress,
  nofootinbib,
  floatfix
]{revtex4}

\usepackage{mathtools}
\usepackage{bm}
\usepackage{dsfont}
\usepackage{graphicx}
\usepackage{pifont}
\usepackage{textcomp}
\usepackage{gensymb}
\usepackage{accents}
\usepackage{upgreek}
\usepackage{array}

\makeatletter
\def\NAT@bibsetnum#1{%
 \setlength{\topsep}{\z@}%
 \NATx@bibsetnum{#1}%
}%

\newcommand*{\supplementarystart}{%
  \close@column@grid%
  \clearpage%
  \onecolumngrid%
  \setcounter{enumiv}{0} 
  \setcounter{equation}{0} 
  \setcounter{figure}{0} 
  \setcounter{table}{0} 
  \setcounter{page}{1}
  \c@secnumdepth=4
  \renewcommand{\theequation}{s\arabic{equation}} 
  \renewcommand{\bibnumfmt}[1]{[s##1]} 
  \renewcommand{\@onlinecite}{s\citealp} 
  \renewcommand{\cite}[1]{{[}\onlinecite{##1}{]}}
  \renewcommand{\thefigure}{s\arabic{figure}}
  \renewcommand{\thetable}{s\Roman{table}}
  \renewcommand{\thepage}{s\arabic{page}}
}
\makeatother

\newcommand{\s}{\sum\limits}

\newcommand{\be}{\begin{equation}}
\newcommand{\e}{\end{equation}}
\newcommand{\beml}{\begin{subequations}}
\newcommand{\eml}{\end{subequations}}
\newcommand{\beq}{\begin{eqnarray}}
\newcommand{\eq}{\end{eqnarray}}
\newcommand{\ba}{\begin{array}}
\newcommand{\ea}{\end{array}}
\newcommand{\bpm}{\begin{pmatrix}}
\newcommand{\epm}{\end{pmatrix}}
\newcommand{\bc}{\begin{cases}}
\newcommand{\ec}{\end{cases}}
\newcommand{\lt}{\left}
\newcommand{\rt}{\right}
\newcommand{\n}{\nonumber}

\newcommand{\ep}{\varepsilon}

\newcommand{\bb}{\boldsymbol}

\DeclareMathOperator{\arctanh}{arctanh}
\DeclareMathOperator{\rot}{rot}

\DeclareMathOperator{\B}{B}

\usepackage{color}

\begin{document}

\title{Plasmon-polariton from a helical state in Dirac magnet}

\author{I.\,V.\,Iorsh}
\affiliation{ITMO University, Saint Petersburg 197101, Russia}

\author{G.\, Rahmanova}
\affiliation{ITMO University, Saint Petersburg 197101, Russia}

\author{M.\,Titov}
\affiliation{Radboud University, Institute for Molecules and Materials, NL-6525 AJ Nijmegen, The Netherlands}
\affiliation{ITMO University, Saint Petersburg 197101, Russia}

\begin{abstract}
Optical field interacting with a topologically protected one-dimensional helical state is shown to support a one-dimensional plasmon-polariton that is characterized by a non-linear dispersion. In a two-dimensional Dirac magnet these electro-optical excitations are confined to domain walls, thus, offering a possibility to manipulate quantum optical states by altering magnetic domain configurations.  An exact spectral equation for such topological plasmon-polariton is derived.    
\end{abstract}

\maketitle

One of the key problems of the modern plasmonics is the search for the new material platforms supporting plasmonic excitations with sufficiently lower losses than in conventional plasmonic materials such as gold or silver{Boltasseva2011}. It was recently realized that the materials exhibiting topologically non-trivial electronic spectrum as well as certain broken symmetries can host extremely low loss plasmonic excitations. Namely, long-lived plasmons and plasmons polaritons have been predicted for a variety of topological and Chern insulators~\cite{Song4658,Kumar2016,Pietro2013,autore2015plasmon,Jin2016}. Moreover, materials with intrinsically broken time reversal symmetry such as Weyl semimetals have been reported to support low-loss plasmonic excitations~\cite{Pellegrini2015,Hofman2016,song2017fermi,Polini2018}.

Dirac magnet is another example of a system with broken time reversal invariance. It  can be realized, e.\,g., in the form of a ferromagnet thin film in a close proximity to a surface of 3D topological insulator or topological semimetal. A perpendicular-to-the plane magnetization component in the ferromagnet induces a finite effective mass of Dirac electrons in the topological insulator.
As the result such a magnetic proximity opens up a band gap in the Dirac electron spectrum, which destroys the two-dimensional Dirac surface state. A one-dimensional domain wall in the ferromagnet is, however, imaged in the Dirac electron system as a zero mass line that supports a helical electronic state. The properties of such a state are similar to those of a quantum Hall edge state. The difference is that the helical state at the domain wall originates in the anomalous Hall effect in the Dirac magnet. 

The physics proposed can be realized e.\,g. in Bi$_2$Se$_3/$EuS interface, where the magnetic proximity effect on topological states has been already experimentally demonstrated \cite{EuS2014,Lee2016}. One can also expect similar phenomena in ZrSiS thin crystal~\cite{Hu2017} that is weakly coupled through an oxide layer to a ferromagnetic thin film. 

\begin{figure}[!h]
\centerline{\includegraphics[width = 0.9\columnwidth]{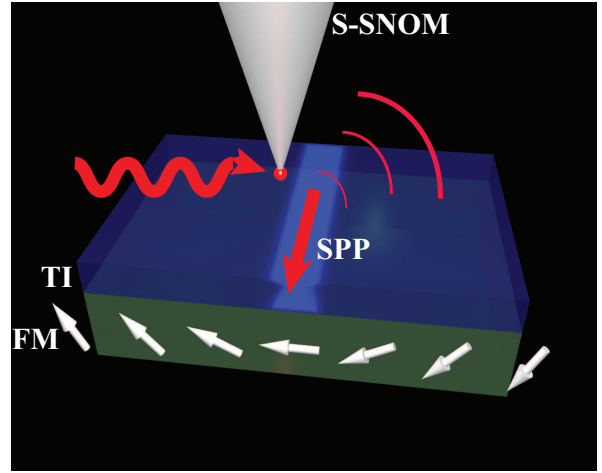}}
\caption{Schematic image of the near-field optical imaging of the ferromagnetic wall via the excitation of the edge plasmon-polariton in Dirac magnet. } 
\label{fig_1}
\end{figure}

In this letter we investigate how these helical electronic states may give rise to one dimensional plasmon-polariton excitations in the presence of THz radiation. Excitation and detection of the plasmon-polaritons are illustrated in Fig.~\ref{fig_1} in a setup that is similar to the one used recently in a series of experiments with scattering SNOM technique~\cite{Chen2012,Fei2012, Basov2016}. 

For a sake of qualitative analysis we choose the simplest model of Dirac magnet that gives rise to helical electronic states 
\be
\label{model}
\mathcal{H}= v\,\lt[\bb{\sigma}\times(\bb{p}-e\bb{A}/c)\rt]_z +\bb{\sigma}\cdot \bb{M}(\bb{r}), 
\e
where the direction $\hat{\bb{z}}$ is chosen perpendicular to the 2D surface, $v\approx 5\,\times 10^5$m$/$s is the effective velocity of Dirac quasiparticles, $\bb{p}$ is the momentum operator, $\bb{A}$ is the vector potential, $e=-|e|$ is the electron charge, $ i\bb{\sigma}$ is the three-dimensional vector of Pauli matrices, and the classical vector field $\bb{M}(\bb{r})$ describes the proximity effect to a ferromagnet. The amplitude of the vector $|\bb{M}|=\Delta$ sets the strength of the exchange coupling between conduction electrons and localized momenta in the ferromagnet.

We consider a flat domain wall that is elongated in $y$ direction and has a characteristic width $a_0$ in $x$ direction. We assume that, away from the domain wall, the magnetization is directed perpendicular to the electron plane $\bb{M}=-\Delta \hat{\bb{z}}$ for $x\ll -a_0$ and $\bb{M}=\Delta \hat{\bb{z}}$ for $x\gg a_0$. The value of exchange coupling $\Delta \approx 4$\,meV defines the spectral gap in the electron spectrum (away from the domain wall). The corresponding length $\ell_\Delta= \hbar v/\Delta \approx 80$\,nm sets the characteristic size of a localized electron wave-function of the helical state in $x$ direction. The domain wall is assumed to be smooth on atomic scales but sufficiently sharp on the scale $\ell_\Delta$, such that $a_0\ll \ell_\Delta$. To avoid optical excitations of bulk states one has to further ensure that the frequency of electromagnetic radiation is smaller than the gap, $\omega\ll \Delta/\hbar \approx 1$\, THz. 

It is evident from Eq.~(\ref{model}) that the terms proportional to $M_x$ and $M_y$ components can be added to the vector potential. Those correspond to additional weak magnetic field in the vicinity of the domain wall that can be neglected. In what follows we simply choose a particular model with $M_x=M_y=0$ and $M_z(x)= \Delta \tanh (x/a_0)$. In the absence of electromagnetic radiation this model gives rise to an exactly solvable spectral problem 
\be
\label{spectral}
v\,\lt[\bb{\sigma}\times\bb{p}\rt]_z\Psi + \Delta \tanh (x/a_0)\,\sigma_z \Psi = \ep \Psi.  
\e
The spectrum of Eq.~(\ref{spectral}) for $|\ep|<\Delta$ is given by a single helical state 
\be
\Psi_{k}(\bb{r}) = \frac{1}{\sqrt{2}} \bpm 1\\1\epm\,F_\nu(x)\, \phi_k(y),
\e
with linear dispersion $\ep_k=\hbar v k$. Here, $\phi_k(y)$ is a plane wave in $y$ direction and $F_\nu(x)$ is a bounded wave-function describing the profile of the helical state in $x$ direction, 
\be
\phi_k(y)=\frac{1}{\sqrt{2\pi}}e^{i k y},\quad F_\nu(x)=\frac{[a_0\B(1/2,\nu)]^{-1/2}}{\cosh^\nu(x/a_0)},
\e
where the parameter $\nu= a_0/\ell_\Delta$ defines the wave-function decay, while $\B(1/2,\nu)=\Gamma(1/2)\Gamma(\nu)/\Gamma(1/2+\nu)$, entering the normalization factor, is the Euler beta function. 

The state $\Psi_{k}$ is evidently an eigenstate of the current operator $\hat{j}_y=e v \sigma_x$ that corresponds to the unidirectional current in $y$ direction (the helical state with the opposite helicity would require $M_z$ changing from positive to negative with increasing $x$). 

In the limit $a_0\ll \ell_\Delta$ one finds a particularly simple expression $F_0(x)=\exp(-|x|/\ell_\Delta)/\sqrt{\ell_\Delta}$ for the helical state profile.

The linear response of the current density $\bb{j}$ to the electric field $\bb{E}$ is defined in the frequency domain by 
\be
\bb{j}(\bb{r},\omega) = \int d^3\bb{r}' \hat{\sigma} (\bb{r},\bb{r}';\omega) \bb{E}(\bb{r}',\omega), \label{jexpr}
\e
where $\hat{\sigma}(\bb{r},\bb{r}';\omega)$ is the conductivity tensor. Due to translational invariance in $y$ direction, the corresponding Fourier transform can be taken in order to express the conductivity tensor in the mixed representation $\hat{\sigma} (x,z;x',z',q;\omega)$, where $q$ is the wave-vector in $y$ direction. 

We shall further assume that the chemical potential of electrons is well within the magnetization induced gap such that $|\mu \pm \hbar \omega| \ll \Delta$, hence the THz radiation with frequency $\omega$ cannot excite electron states in the bulk. In this regime, the optical conductivity can be expressed by Kubo formula~\cite{Haug2009} that takes into account only the helical electron state, while the only non-vanishing component of the conductivity tensor is given by
\begin{align}
\sigma_{yy}&(x,z;x',z',q;\omega)=\frac{ie^2}{\hbar}v\; \delta(z)\delta(z')\, \frac{F_\nu^2(x)F_\nu^2(x')}{q(\omega-v\,q)}\n\\
&\times \int_{-\infty}^{\infty} \frac{dk}{2\pi} \,\lt[f(\ep_{k-q/2})-f(\ep_{k+q/2})\rt],
\label{Kubo}
\end{align}
where $f(\ep)=\lt[e^{(\ep_k-\mu)/T}+1\rt]^{-1}$ is the Fermi distribution function. For $T\ll \Delta$, the integration in Eq.~(\ref{Kubo}) is readily performed with the result
\be
\sigma_{yy}=\bar{\sigma}_{\omega,q} \;F_\nu^2(x)F_\nu^2(x')\; \delta(z)\delta(z'),
\e 
where we have defined a ``one-dimensional'' conductivity
\be
\label{sigma1D}
\bar{\sigma}_{\omega,q}=-\frac{1}{2\pi i}\;\frac{\alpha\, v}{k_0- q\,v/c}, \qquad k_0=\omega/c,
\e
with $\alpha = e^2/\hbar c \approx 1/137$ standing for the fine structure constant.

The result of Eq.~(\ref{sigma1D}) describes a one-dimensional plasmon with the trivial linear dispersion $\omega_q = v q$. It is known in the theory of 1D electronic systems~\cite{Carmelo1996,Ng1997,Castellani1999} that, in contrast to single electron excitations, such a plasmon may remain a good quasi-particle even in the presence of interactions.

When THz radiation is applied to the system as shown in Fig.~\ref{fig_1}, the plasmon is transformed to another quasiparticle that is called plasmon-polariton. To find its dispersion it is necessary to solve the Maxwell equation
\begin{align}
\rot\rot \bb{E}(\bb{r},\omega)=k_0^2 \bb{E}(\bb{r},\omega)-\frac{4i\pi k_0}{c}\bb{j}(\bb{r},\omega),
\end{align}
with the electric current $\bb{j}$ defined by Eq.~\eqref{jexpr}. This equation is equivalent to the integral equation of the form,
\begin{align}
E_{\gamma}(\bb{r},\omega)=&i\frac{4\pi k_0}{c}\s_{\alpha\beta} \int d^{3}\bb{r}'\; G_{\gamma\alpha}(\bb{r}-\bb{r}';\omega)\n \\
&\times\int d^3\bb{r}''\sigma_{\alpha\beta}(\bb{r}',\bb{r}'';\omega)E_{\beta}(\bb{r}'',\omega),
\label{Eexpr}
\end{align}
where $G_{\alpha\beta}(\bb{r}-\bb{r}';\omega)$ stands for the dyadic Green's function in vacuum. 

The Equation~(\ref{Eexpr}) is non-trivial only for the component $E_y$. Taking into account again the translational invariance in $y$ direction, we obtain
\begin{align}
E_{y}&(x,z;q;\omega)=
i\bar{\sigma}_{\omega,q} \frac{4\pi k_0}{c} \int dx'\; G_{yy}(x-x',z;q)\n \\
&\times F^2_\nu(x')\int dx''F^2_\nu(x'') E_y (x'',0;q;\omega),
\label{Eexpr1}
\end{align}
where we introduce the Fourier transformed Green's function $G_{yy}(x-x',z-z';q)$ in the mixed representation. 

To obtain the spectral equation on plasmon-polariton we multiply Eq.~\eqref{Eexpr1} by $F_\nu^2(x)\delta(z)$ and integrate over $x$ and $z$. The procedure leads to the relation
\be
\label{spectral1}
1 = i\bar{\sigma}_{\omega,q} \frac{4\pi k_0}{c} \int\!\!\!\int dx\,dx'\; F^2_\nu(x)G_{yy}(x-x')F^2_\nu(x'),
\e
where $G_{yy}(x-x')=G_{yy}(x-x',0;q)$. This Green's function can be represented as
\be
\label{resG}
G_{yy}(x-x')= \frac{1}{2k_0^2}\int \frac{dk}{2\pi} \frac{k_0^2-q^2}{\sqrt{k^2+q^2-k_0^2}} e^{ik(x-x')}.
\e
Substituting Eq.~(\ref{resG}) in Eq.~(\ref{spectral1}) we perform the integrations over $x$ and $x'$ analytically with the result
\be
\label{disp}
1= \frac{i\bar{\sigma}_{\omega,q}}{\omega} \frac{k_0^2-q^2}{\Gamma^4(\nu)} \int_{-\infty}^\infty du \frac{\lt|\Gamma(\nu(1+iu/2))\rt|^4}{\sqrt{u^2-\ell_\Delta^2(k_0^2-q^2)}}, 
\e
that defines the exact dispersion relation $\omega_q$ for the surface plasmon-polariton.

\begin{figure}[!h]
\centerline{\includegraphics[width = 0.9\columnwidth]{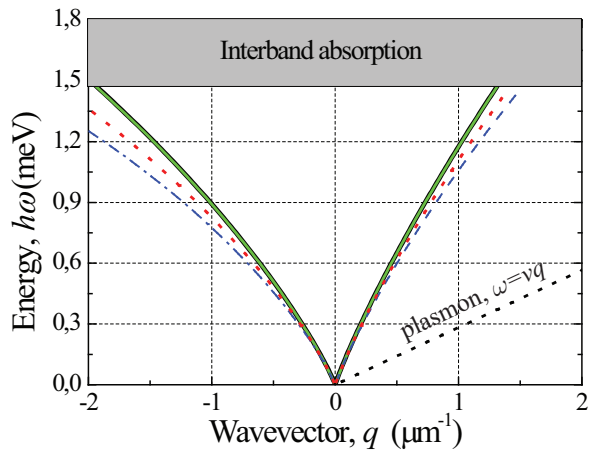}}
\caption{Dispersion of the one-dimensional topological plasmon polariton for characteristic material parameters of FM/ZrSiS heterostructure, where $q$ is the wave vector component along the electron helical state. For $q\ll 1/\ell_{\Delta} \approx 5$\,$\mu$m$^{-1}$, the two branches of the dispersion are nearly symmetric and are given by Eq.~(\ref{limit}). For $q\gtrsim 1/\ell_{\Delta}$ one observes a strong asymmetry of the branches. Different lines correspond to different choices of the domain wall width $a_0=\nu \ell_\Delta$:  black line ($\nu=0$), green line ($\nu=0.05$), dotted red ($\nu=1$) and dashed blue ($\nu=2$). Other parameters of the calculation are listed in the text.}
\label{fig_2}
\end{figure}

It is instructive to analyze the dispersion relation defined by Eq.~(\ref{disp}) in the limit $\nu= a_0/\ell_\Delta \ll 1$. For $\nu=0$ Eq.~(\ref{disp}) is simplified to 
\be
\label{eq_simplified}
1-\tilde{v}\tilde{q}=\alpha \tilde{v}(\tilde{q}^2-1)\mathcal{S}(\kappa)/2\pi,
\e
where $\tilde{q}=q c/\omega$, $\tilde{v}=v/c$, $\kappa^2=1+\ell_\Delta^2(k_0^2 - q^2)/4$ and
\begin{align}
\mathcal{S}(\kappa)=\, &\int_{-\infty}^\infty \!\!\! du\;(1+u^2)^{-2}(1+u^2-\kappa^2)^{-1/2}\n\\
&= \kappa^{-3}\lt[(1+\kappa^2)\arctanh(\kappa)-\kappa\rt]. 
\end{align}
Furthermore, we note that the condition $\hbar\omega \ll \Delta c/v$ is always fulfilled, consequently $\ell_\Delta k_0 \ll 1$. Thus, for sufficiently small momenta $q \ll \ell^{-1}_\Delta$ we can always regard the parameter $1-\kappa^2$ as small and positive. In this limit we can simply rewrite Eq.~(\ref{eq_simplified}) as
\be
\label{eq_simplified1}
1-\tilde{v}\tilde{q}=\frac{\alpha}{2\pi} \tilde{v}(\tilde{q}^2-1)\lt[\ln\lt(\frac{16}{\ell_\Delta^2(q^2-k_0^2)}\rt) -1\rt].
\e
This equation can be explicitly solved in the limit $\omega \ll q c$ (or $\tilde{q}\gg 1$) with the result
\be
\label{eq_simplified2}
\omega^\pm_{q}= \frac{q v}{2}\lt(1\pm \sqrt{1+\frac{2\alpha c}{\pi v}\lt(\ln\frac{16}{q^2\ell_\Delta^2} -1\rt)}\rt).
\e
Note that we formally restore the plasmon dispersion for $\alpha=0$. Taking the leading logarithm approximation in Eq.~(\ref{eq_simplified2}) we obtain a particularly simple relation
\be
\label{limit}
\omega^\pm_q = \pm qv\sqrt{\frac{\alpha c}{\pi v} \ln\frac{1}{|q| \ell_\Delta}},\qquad  q\ll 1/\ell_\Delta,
\e
where the branches $\omega^+$ and $\omega^-$ correspond to positive and negative values of the momentum $q$, respectively. The symmetry between the branches is, however, broken for $q$ approaching $1/\ell_\Delta$. To describe the dispersion relation at sufficiently large values of momenta one still has to refer to Eqs.~(\ref{disp}) or (\ref{eq_simplified}).

In Fig.~\ref{fig_2} we illustrate the dispersion relation obtained numerically from Eqs.~(\ref{disp}) and (\ref{eq_simplified}) for realistic parameter values. We find that for $q> 1/\ell_\Delta$ the dispersion differs substantially from the one for Hall effect edge magneto-plasmon polariton~\cite{Volkov1988}. 

For realistic analysis one still has to take into account the presence of the substrate by replacing the dyadic Green's function in Eq.~(\ref{spectral}) with the one for a particular heterostructure. The latter can be routinely calculated for an arbitrary layered system~\cite{Tomas1995}. Here, we still refer to  Eqs.~(\ref{disp}) and (\ref{eq_simplified}) with the parameters that are characteristic for a surface of ZrSiS~\cite{Hu2017} in proximity to a generic ferromagnetic film with a saturation magnetization of 0.5\,T.  In such a setup we have $\Delta\approx 1.5$\,meV and $v\approx 4.3\times 10^5$\,m$/$s, consequently, we find $\ell_\Delta\approx 0.2$\,$\mu$m.  

In Fig.~\ref{fig_2} we plot the dispersion of the plasmon polariton defined by Eq.~(\ref{disp}) for different values of $a_0=\nu \ell_\Delta$. 
A typical domain wall with the width $a_0\approx 10$\,nm corresponds to $\nu\approx 0.05$. It can be seen that the dispersion for the case $\nu=0.05$ does agree very well with the one given by Eq.~(\ref{eq_simplified}) for $\nu=0$. The discrepancies between the results of Eqs.~(\ref{disp}) and (\ref{eq_simplified}) become visible only for $\nu\approx 1$. For small values of momenta the results are well described by the limiting expression of Eq.~(\ref{limit}). We note, that our choice of parameters correspond to $\alpha c/\pi v \approx 1.6$.

\begin{figure}[!h]
\centerline{\includegraphics[width = 0.9\columnwidth]{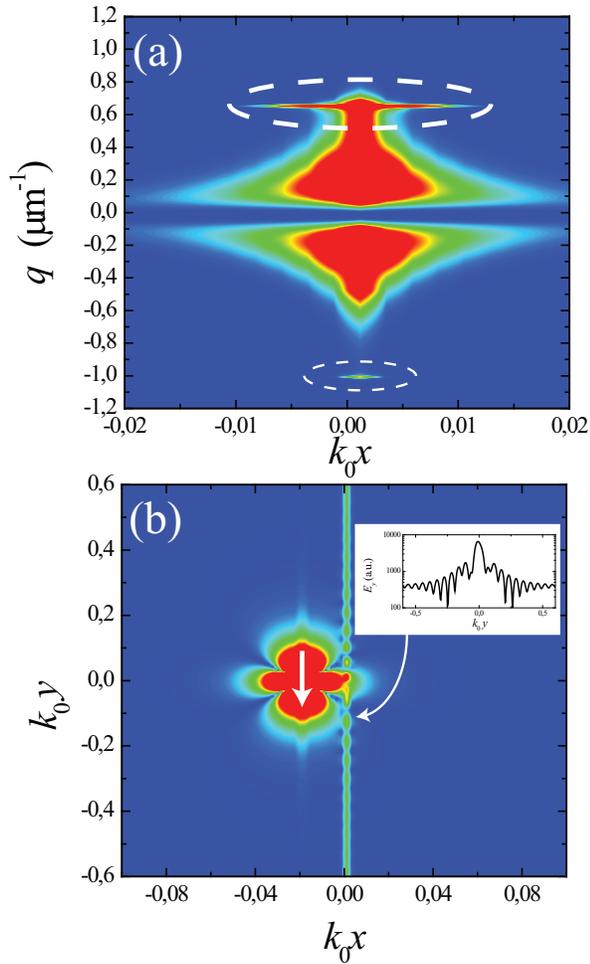}}
\caption{(a) Fourier map of the scattered electric field from a dipole positioned at $\bb{r}_0=(-5,0,1)$\,$\mu$m assuming the dipole frequency $\hbar\omega=0.75$\,meV.
The dashed regions show the excitation of the surface plasmon modes. (b) Real space map of the $y$ component of the total electric field. Inset shows the field profile at $x=0$.} 
\label{fig_3}
\end{figure}

Let us now consider a particular scenario for the excitation of the helical plasmon-polariton that has been already undertaken in several experiments \cite{}. Such an excitaiton can be achieved by placing a point-dipole at the position $\bb{r}_0=(x_0,0,z_0)$ in a vicinity of the domain wall. We assume that the dipole is oriented along $y$ axis, hence the $y$ component of electric field is given by 
\begin{align}
&E_y(x,z,q;\omega)=G_{yy}(x-x_0,z-z_0;q)+\n \\
&+\alpha \tilde{v}\,\Lambda(q,x_0,z_0)\int dx' G_{yy}(x-x',z;q)F_{\nu}^2(x'), 
\label{EscatF}
\end{align}
where we again adopt the mixed representation of the dyadic Green's function used in Eq.~(\ref{Eexpr1}), and 
\begin{align}
\Lambda(q,x_0,z_0)=
\frac{\int dx'' G_{yy}(x''-x_0,-z_0;q)F_{\nu}^2(x'')}{1-\tilde{v}\tilde{q}-\alpha \tilde{v}(\tilde{q}^2-1) \mathcal{S}(\kappa)/2\pi}.
\end{align}

In Fig.~\ref{fig_3}(a) we plot the second term at the right hand side of Eq.~\eqref{EscatF} that represents the scattered radiation. At the plot we choose $\hbar\omega=0.75$\, meV and  $\bb{r}_0=(-5,0,1)$\,$\mu$m. 

The plot shows that, besides the broad distribution of the near field, there exist two narrow peaks corresponding to the excitation of the surface plasmon polaritons (marked with dashed white ellipses). Indeed, the central positions of these peaks match the dispersion relation of Eqs.~(\ref{disp},\ref{eq_simplified1}) (see Fig.~\ref{fig_2})  
at the corresponding frequencies. 

It also worth noting that both the peak positions and the peak intensity are somewhat different for positive and negative $q$ (for $|q|\gtrsim 1/\ell_\Delta$) highlighting the helicity of the corresponding electron state. In Fig.~\ref{fig_3}(b) we show the real space map of the total electric field given by Eq.~\eqref{EscatF}. The map illustrates the point dipole exciting the edge plasmon polariton. Moreover, as can be seen in the inset, the interference between the incident field and the excited plasmon-polariton leads to the characteristic oscillations of the absolute value of the field. The spatial period of the oscillations can be estimated as $\pi/(k_0-q_{P})$, where $q_{P}$ is the plasmon-polariton wavevector. Thus, the dispersion of the plasmon-polariton can be directly probed directly in the near field scanning optical microscopy measurements. 

In conclusion we considered a model of a helical electron state that can be formed along at a surface of 3D topological insulator or topological semimetal that is brought to a proximity to a ferromagnet thin film with a flat domain wall. Such a state is analogous to the quantum Hall edge state but arise due to the anomalous Hall effect. We demonstrated that the state supports edge plasmon-polariton excitations and computed, for a simple model, the plasmon-polariton dispersion relation. We discuss a possibility to observe the effect with THz near field scanning optical microscopy. 

{\it Acknowledgments} --- M.T. acknowledges the support from the Russian Science Foundation under Project 17-12-01359 and the support from the JTC-FLAGERA Project GRANSPORT.

\bibliography{TopMagn}

\end{document}